\documentclass[aps,prb,twocolumn,showpacs,floatfix]{revtex4}
\usepackage{graphicx}
\begin{document}

\title{Simple implementation of complex functionals: scaled
selfconsistency}

\author{Matheus P. Lima, Luana S. Pedroza, Antonio J. R. da Silva and A. Fazzio}
\affiliation{Instituto de F\'{\i}sica, Universidade de S\~ao Paulo, 
S\~ao Paulo, Brazil}

\author{Daniel Vieira, Henrique J. P. Freire and K. Capelle}
\email{capelle@ifsc.usp.br}
\affiliation{Departamento de F\'{\i}sica e Inform\'atica,
Instituto de F\'{\i}sica de S\~ao Carlos,
Universidade de S\~ao Paulo,
Caixa Postal 369, 13560-970 S\~ao Carlos, SP, Brazil}

\date{\today}

\begin{abstract}
We explore and compare three approximate schemes allowing simple implementation
of complex density functionals by making use of selfconsistent 
implementation of simpler functionals: (i) post-LDA evaluation of complex 
functionals at the LDA densities (or those of other simple functionals); 
(ii) application of a global scaling factor to the potential of the
simple functional; and (iii) application of a local scaling factor to that
potential. Option (i) is a common choice in density-functional calculations.
Option (ii) was recently proposed by Cafiero and Gonzalez.
We here put their proposal on a more rigorous basis, by deriving it, and
explaining why it works, directly from the theorems of density-functional
theory. Option (iii) is proposed here for the first time. We provide 
detailed comparisons of the three approaches among each other and with fully
selfconsistent implementations for Hartree, local-density, 
generalized-gradient, self-interaction corrected, and meta-generalized-gradient
approximations, for atoms, ions, quantum wells and model Hamiltonians. 
Scaled approaches turn out to be, on average, better than post-approaches, 
and unlike these also provide corrections to eigenvalues and orbitals.
Scaled selfconsistency thus opens the possibility of efficient and reliable 
implementation of density functionals of hitherto unprecedented complexity.
\end{abstract}


\maketitle

\newcommand{\be}{\begin{equation}}
\newcommand{\ee}{\end{equation}}
\newcommand{\bea}{\begin{eqnarray}}
\newcommand{\eea}{\end{eqnarray}}
\newcommand{\bi}{\bibitem}
\newcommand{\la}{\langle}
\newcommand{\ra}{\rangle}
\newcommand{\ua}{\uparrow}
\newcommand{\da}{\downarrow}
\renewcommand{\r}{({\bf r})}
\newcommand{\rp}{({\bf r'})}
\newcommand{\rpp}{({\bf r''})}

\section{Introduction}
\label{intro}

Density-functional theory \cite{kohnrmp,dftbook,parryang} is the driving
force behind much of todays progress in electronic-structure calculations.
Progress in density-functional theory (DFT) itself depends on the twin
development of ever more precise density functionals and of ever more
efficient computational implementations of these functionals. The first line 
of development, functionals, has lead from the local-density approximation
(LDA) to generalized-gradient approximations (GGAs), hybrid functionals,
and on to meta-GGAs and other fully nonlocal approximations, such 
as exact-exchange (EXX) and self-interaction corrections (SICs).\cite{perdewreview,oepreview}

As functionals get more and more complex, the second task, implementation,
gets harder and harder. Indeed, very few truly selfconsistent implementations
of beyond-GGA functionals exist, and even GGAs are still sometimes 
implemented non-selfconsistently. At the heart of the problem is not
so much the actual coding (although that can also be a formidable task,
considering the complexity of, e.g., meta-GGAs), but rather obtaining the
exchange-correlation ($xc$) potential $v_{xc}\r$ corresponding to a given 
approximation to the $xc$ energy $E_{xc}[n]$. Hybrid functionals, meta-GGAs,
EXX and SICs are all orbital functionals, {\em i.e.}, functionals of the
general form $E^{orb}_{xc}[\{\varphi_i[n]\}]$, where our notation indicates 
an explicit dependence on the set of Kohn-Sham (KS) orbitals $\varphi_i\r$. 
This set may be restricted to the occupied orbitals, but may also include 
unoccupied orbitals. Since by virtue of the Hohenberg-Kohn (HK) theorem 
these orbitals themselves are density functionals, any explicit orbital 
functional is an implicit density functional. 

The HK theorem itself, however, does not provide any clue how to obtain
the $xc$ potentials, i.e., how to calculate the functional derivative
\be
v_{xc}[n]\r= \frac{\delta E_{xc}^{orb}[\{\varphi_i\}]}{\delta n\r}
\ee
of a functional whose density dependence is not known explicitly. Three
different solutions to this dilemma have been advanced in the literature.

(i) The formally correct way to implement an implicit density functional in
DFT is the optimized-effective potential (OEP) algorithm\cite{oepreview}
[also known as the optimized potential method (OPM)], which results in an 
integral equation for the KS potential corresponding to the orbital functional.
The first step of the derivation of the OEP is to write
\bea
v_{xc}[n]\r= \frac{\delta E_{xc}^{orb}[\{\varphi_i\}]}{\delta n\r}
\label{oepinteg1}
\\
=\int d^3r' \sum_i
\left[\frac{\delta E_{xc}^{orb}[\{\varphi_i\}]}{\delta \varphi_i\rp}
\frac{\delta \varphi_i\rp}{\delta n\r}
+ c.c. \right]
\label{oepinteg2}
\\
=\int d^3r' d^3r'' \sum_i
\left[\frac{\delta E_{xc}^{orb}[\{\varphi_i\}]}{\delta \varphi_i\rp}
\frac{\delta \varphi_i\rp}{\delta v_s\rpp}\frac{\delta v_s\rpp}{\delta n\r}
+ c.c. \right].
\label{oepinteg3}
\eea
Further evaluation of Eq.~(\ref{oepinteg3}) gives rise to an integral equation
that determines the $v_{xc}[n]$ belonging to the chosen orbital functional
$E^{orb}_{xc}[\{\varphi_i[n]\}]$.
This OEP integral equation must be solved at every step of the selfconsistency
cycle, which considerably increases demands on memory and computing time.
Often (in particular for systems that do not have spherical symmetry) the
integral equation is simplified by means of the Krieger-Li-Iafrate (KLI) 
approximation,\cite{kli} but even within this approximation, or other recently 
proposed simplifications,\cite{kuemmelperdew,wuyang} an OEP calculation is 
computationally more expensive than a traditional KS calculation. 
A separate issue is the unexpected behaviour of the OEP in finite basis sets, 
which was recently reported to cast doubt on the reliability of the 
method.\cite{staroverov} The OEP prescription moreover
assumes in the first step that the {\em orbital} derivative 
$\delta E_{xc}^{orb}[\{\varphi_i\}]/\delta \varphi_i\rp$ can 
be obtained, and even that is not a trivial task for complex orbital-dependent
functionals such as meta-GGA and hyper-GGA. (For the Fock term, on the 
other hand, this derivative is simple, and the implementation of the Fock 
term by means of the OEP is becoming popular under the name exact exchange 
[EXX]). 

(ii) Instead of taking the variation of $E_{xc}[\{\varphi_i[n]\}]$ with 
respect to $n\r$, as in the OEP, one often makes $E_{xc}$ stationary with 
respect to the orbitals. When applied to the Fock term, this is simply the 
Hartree-Fock (HF) method, but similar procedures are commonly applied within 
DFT to hybrid functionals, such as B3-LYP. Selfconsistent implementations of 
meta-GGA reported until today are also selfconsistent with respect to the 
orbitals, not the densities.\cite{handymgga,mggatests1,mggatests2} 
For the Fock term, one finds empirically that the occupied orbitals obtained 
from orbital selfconsistency are quite similar to those obtained from density 
selfconsistency, but such empirical finding is no guarantee that the similarity
will persist for all possible functionals or systems --- and in any case it 
does not extend to the unoccupied orbitals. Computationally, the resulting
nonlocal HF-like potential makes the solution of the Kohn-Sham equation more 
complex than the local potential resulting from the OEP (although the 
potential itself is obtained in a simpler way). Moreover, as the OEP, this 
prescription also requires that the orbital derivative
$\delta E_{xc}^{orb}[\{\varphi_i\}]/\delta \varphi_i\rp$ can be obtained, 
which need not be simple for complex functionals.

(iii) Less demanding than density-selfconsistency or orbital-selfconsistency 
are post-selfconsistent implementations, in which a selfconsistent KS 
calculation is performed with a simpler functional, and the resulting orbitals 
and densities are once substituted in the more complex one. This strategy is 
sometimes used to implement GGAs in a post-LDA way, and has been applied
also to meta-GGAs. It is much simpler than the other two possibilities,
but does not lead to KS potentials, orbitals, and single-particle energies 
associated with the complex functional. It simply yields the total energy the 
complex functional produces on the densities obtained from the simpler 
functional.

As this brief summary shows, each of the three choices has its own distinct
set of advantages and disadvantages. A method that is simple to implement,
reliable, and provides access to total energies as well as single-particle 
quantities would be a most useful addition to the arsenal of computational DFT. 

A method we consider to have the desired characteristics was recently proposed 
by Cafiero and Gonzalez,\cite{cafiero} and consists in the application of 
a scaling factor to the $xc$ potential of a simple functional (which can
be implemented fully selfconsistently), such that it approximates the 
potential expected to arise from a complex functional (which need not be
implemented selfconsistently). This fourth possibility, which we call 
{\em globally-scaled selfconsistency} (GSSC), is analysed in Sec. \ref{ssc} 
of the present paper.

We believe the derivation presented by Cafiero and Gonzalez to lack 
rigour at a crucial step, and thus dedicate Sec.~\ref{cg} to an 
investigation of their procedure. Our investigation leads to an alternative 
derivation of the same approach, described in Sec.~\ref{alternative}. Next,
we provide, in Sec. \ref{validity}, an analysis of the validity of the 
scaling approximation, arriving at the counterintuitive (but numerically 
confirmed and explainable) conclusion that the success of this approximation 
is not due to the smallness of the neglected term relative to the one kept.
We also suggest a modification of the approach, which we call 
{\em locally-scaled selfconsistency} (LSSC) and describe in 
Sec.~\ref{localssc}.

In Secs.~\ref{atomtests} and \ref{modeltests} we provide extensive numerical 
tests of GSSC and LSSC, 
and also of the more common post-selfconsistent mode of implementation [here 
denoted P-SC, option (iii) above]. Note that although P-SC is quite commonly
used, it has not been systematically tested in situations where the exact 
result or the fully selfconsistent result is known. All three schemes are 
applied to atoms and ions in Secs.~\ref{atomsionsE} and \ref{atomsionseps},  
to one-dimensional Hubbard chains, in Sec.~\ref{hubbard}, 
and to semiconductor quantum wells, in Sec.~\ref{qwell}. We compare
Hartree, LDA, GGA, MGGA and SIC calculations, implemented via P-SC, 
GSSC and LSSC, among each other, and, where possible, with results obtained 
from fully selfconsistent implementations of the same functionals. 

Our main conclusion is that scaled selfconsistency works surprisingly well, 
for all different combinations of systems and functionals we tried. Scaled 
selfconsistency yields better energies than post-selfconsistent 
implementations, and moreover provides access to orbitals and eigenvalues.
This conclusion opens the possibility of efficient and reliable implementation
of density functionals of hitherto unprecedented complexity.

\section{Scaled selfconsistency}
\label{ssc}

In this section we first briefly review the approach suggested by Cafiero and 
Gonzalez (CG),\cite{cafiero} and point out where we believe their development 
to lack rigour. Next we present an alternative derivation and suggest a
modification. 

\subsection{The proposal of Cafiero and Gonzalez}
\label{cg}

CG consider a complex functional, $E_{xc}^B[n]$ with potential $v_{xc}^B\r$, 
and a simple functional, $E_{xc}^A[n]$ with potential $v_{xc}^A\r$. 
They then define the energy ratio 
\be
F_{xc}[n]=\frac{E_{xc}^B[n]}{E_{xc}^A[n]}
\ee
and differentiate $E_{xc}^B[n]=F_{xc}[n] E_{xc}^A[n]$ with respect to the
density $n\r$, obtaining, in their notation,
\be
v_{xc}^B\r=F_{xc}[n]v_{xc}^A\r 
+ E_{xc}^A[n] \frac{\partial F_{xc}[n]}{\partial n\r}.
\label{wrong}
\ee
Next, CG argue that at the selfconsistent density, $n^*$, 
$E_{xc}^B[n^*]=E_{xc}^A[n^*]$, so that $F_{xc}[n^*]\to 1$. According to CG, 
the last term in Eq.~(\ref{wrong}) is then equal to zero. Under this 
'constraint'\cite{cafiero} they obtain
\be
v_{xc}^B\r=F_{xc}[n]v_{xc}^A\r
\label{cgscaling}
\ee
which implies that an (approximate) selfconsistent implementation of
the complicated functional $E_{xc}^B[n]$ can be achieved by means of a
selfconsistent implementation of the simple functional $E_{xc}^A[n]$,
if at every step of the selfconsistency cycle the potential corresponding
to functional A is multiplied by the ratio of the energies of B and A.
This procedure completely avoids the need to ever calculate or implement
the potential corresponding to $B$. CG go on to test their proposal for
a few functionals and systems, and claim very good numerical agreement 
between approximate results obtained from (\ref{cgscaling}) and fully 
selfconsistent or post-selfconsistent implementations.

We were initially quite surprised by these good results, as
the preceding argument lacks rigour at a key step. Apart from the 
use of partial derivatives instead of variational ones (which is probably
just a question of notation), we do not see why at the selfconsistent
density the simple and the complex functional must approach the same value,
i.e., why $F[n^*]$ should approach one. Moreover, whatever value $F$ 
approaches, this value should not be used to simplify the equation for
the $xc$ potential because the derivative must be carried out {\em before}
substituting numerical values in the energy functional, not afterwards. 

\subsection{An alternative derivation} 
\label{alternative}

In view of these troublesome features of the original derivation we below
present an alternative rationale for the same approach.
We start by writing the same identity the CG approach is based on,
\be
E^B[n]=\frac{E^B[n]}{E^A[n]}E^A[n] =: F[n] E^A[n].
\ee
Note that we write this identity for an arbitrary energy functional, as
the approach is, in principle, not limited to $xc$ functionals. Indeed, 
below we will show one application in which we deal with the entire 
interaction energy (Hartree + $xc$). Next, we take the functional derivative 
of the product $F[n] E^A[n]$,
\bea
v^B\r \equiv  \frac{\delta E^B[n]}{\delta n\r}
= \frac{\delta F[n]}{\delta n\r} E^A[n]
+ F[n]\frac{\delta E^A[n]}{\delta n\r}
\\
= \frac{\delta F[n]}{\delta n\r}E^A + F[n] v^A\r.
\eea
The functional derivative of the ratio $F[n]$ is
\be
\frac{\delta }{\delta n\r}\frac{E^B}{E^A}=
\frac{E^A[n] v^B\r - E^B[n] v^A\r}{E^A[n]^2}.
\ee
Substitution of this in the previous equation yields the identity
\be
v^B\r = \frac{E^A[n] v^B\r - E^B[n] v^A\r}{E^A[n]} + F[n] v^A\r.
\label{identity}
\ee
Neglecting the first term leaves us with the simple expression
\be
v^B\r \approx v^{GSSC}\r := F[n] v^A\r
\label{gssc}
\ee
as an approximation to $v^B$. In the following this approximation is called
{\em globally scaled selfconsistency} (GSSC). {\em Scaling} here refers 
to the multiplication of the simple potential $v^A\r$ by the energy ratio 
$F[n]$ in order to simulate the more complex potential $v^B\r$, and the 
specifier {\em globally} is employed in anticipation of a local variant
developed in Sec. \ref{localssc}.

\subsection{Validity of scaled selfconsistency}
\label{validity}

The form of Eq.(\ref{identity}) suggests a simple explanation of why 
scaled selfconsistency can lead to results close to those obtained from 
a fully selfconsistent implementation of functional B: 
Clearly, $v^{GSSC}\r$ is a good approximation to $v^B\r$ if the
first term in (\ref{identity}) can be neglected compared to the second.
Globally scaled selfconsistency is thus guaranteed to be valid if the
validity criterium
\be
C_1\r:=\frac{\left|E^A[n] v^B\r - E^B[n] v^A\r\right|}{|E^B[n] v^A\r|}<<1
\label{criterium1}
\ee
is satisfied at all points in space.
Note that this is conceptually a very different requirement from
$F[n^*]\to 1$, although it leads to the same final result.

Interestingly, criterium (\ref{criterium1}) is a sufficient, but not at all a
necessary, criterium for applicability of the GSSC approach. In fact, in the 
applications reported below we have frequently encountered situations in which 
the first term on the right-hand side in Eq.~(\ref{identity}) is not much 
smaller, but comparable to, or even larger than, the second term. The 
empirical success of scaled selfconsistency, claimed in Ref.
\onlinecite{cafiero} and confirmed below for a wide variety of functionals 
and systems, must thus have another explanation, rooted in systematic error 
cancellation.

One source of error cancellation is the fact that
Eq. (\ref{criterium1}) is a point-wise equation, which, in principle,
must be satisfied for every value of ${\bf r}$. For the calculation of
integrated quantities (such as total energies) or quantities determined by
the values of $v^B\r$ at all points in space (such as KS eigenvalues) a
violation of (\ref{criterium1}) at some point in space can be
compensated by that at some other point. We have numerically verified
that this indeed happens: the sign of the term neglected in GSSC can be
different in different regions of space, reducing the integrated error.

The extent to which errors at different points cancel in the 
application of the scaling factor can be estimated by replacing 
Eq. (\ref{criterium1}) by
\be
C_2:=
\frac{\int d^3r\, E^A[n] v^B\r-E^B[n] v^A\r}{\int d^3r E^B[n] v^A\r}<<1.
\label{criterium2}
\ee
Representative values of $C_2$ are also reported below.

In applications we have frequently obtained excellent total energies 
even in situations in which this criterium also fails.
A second source of error cancellation is the DFT total-energy expression
\be
E_0[n]=\sum_i\epsilon_i - E_H[n] -\int d^3r \, n\r v_{xc}\r + E_{xc}[n],
\label{etot}
\ee
where, ideally, $v_{xc}$ is the functional derivative of $E_{xc}[n]$.
Scaled selfconsistency is applied in situations in which $E_{xc}[n]$ 
is known, but $v_{xc}$ is hard to obtain or to implement. Hence it 
replaces $v_{xc}$ by $v_{xc}^{GSSC}$ in the second-to-last term, but
the last term is evaluated using the original $E_{xc}$. The entire
expression is evaluated on the selfconsistent density arising from a
Kohn-Sham calculation with potential $v_{xc}^{GSSC}$. In situations in 
which all terms in (\ref{etot}) can be obtained exactly, we have verified 
numerically that the errors arising from different terms in the GSSC total
energy make contributions of similar size but opposite sign to the 
total error. Representative analyses of this type are given below.

Numerical comparison of the exact equation (\ref{identity}) and its 
approximation (\ref{gssc}) with results obtained from the local 
(\ref{criterium1}) or the integrated (\ref{criterium2}) validity criterium,
and with the errors of each term in the total-energy expression (\ref{etot}),
permits us to identify the reason for the success of GSSC. Anticipating 
results presented in detail in later sections, we find that
the excellent approximations to total energies obtained from the simple
scaling approach are to a large extent due to the error cancellation
between different terms of the total energy, and depend only very weakly
on the smallness of the neglected term relative to the one kept.

We end this discussion of validity and errors by pointing out that although 
frequently a violation of the criteria (\ref{criterium1}) or
(\ref{criterium2}) still provides good energies, we have never observed the 
opposite situation, i.e., bad energies in situations in which 
(\ref{criterium1}) or (\ref{criterium2}) are satisfied. 

\subsection{Locally-scaled selfconsistency}
\label{localssc}

As a result of the previous section, we have the expression
\bea
v^B[n]({\bf r}) \approx
\nonumber \\
v^{GSSC}[n]({\bf r}) =
\frac{E^B[n]}{E^A[n]}v^A[n]({\bf r})
=F[n]v^A[n]({\bf r})
\eea
for globally-scaled selfconsistency. Note that the scaling factor depends
on the density, i.e., is updated in every iteration of the selfconsistency
cycle, but does not depend on the spatial coordinate ${\bf r}$, i.e., is the
same for all points in space.

Clearly,  such a global scaling factor cannot properly account for the
fine point-by-point differences expected between the potentials 
$v^A\r$ and $v^B\r$. In an attempt to account for these differences, we 
introduce a local scaling factor, based on the observation that 
many functionals (in particular, LDA, GGA and MGGA) can be cast in the form
\be
E[n] = \int d^3r\, e(n(r)),
\ee
where the {\em energy density} $e(n)$ is defined (up to a
total divergence) by the above expression, and the dependence of
$e(n)$ on $n\r$ may be both explicit (as in LDA) or implicit (as in
meta-GGA). The energy density allows us to introduce a {\em local
scaling factor} $f[n]({\bf r})$, according to
\bea
v^B[n]({\bf r}) \approx
v^{LSSC}[n]({\bf r}) =
\\
\frac{e^B[n]({\bf r})}{e^A[n]({\bf r})}v^A[n]({\bf r})
=:f[n]({\bf r})v^A[n]({\bf r}).
\eea
In the applications below we refer to $v^{GSSC}[n]$ as the globally scaled
potential (with  upper-case scaling factor $F$), and to $v^{LSSC}$ as locally 
scaled potential (with lower-case scaling factor $f\r$). 

We also explored a variety of alternative scaling schemes, such as, e.g., 
application of scaling factors to the full effective potential instead of 
just its $xc$ contribution, or directly to the density. None of these 
consistently led to better results than the GSSC and LSCC procedures, and
we refrain from presenting detailed results from alternative schemes here.

\section{Tests and applications to atoms and ions}
\label{atomtests}

In this section we compare post-selfconsistent implementations and 
globally and locally scaled implementations of different density functionals 
for total energies and Kohn-Sham eigenvalues of atoms and ions.
We provide comparisons of these modes of implementation among each 
other and, where possible, with fully selfconsistent applications. We also 
provide representative evaluations of the validity criteria discussed in 
Sec. \ref{validity}.

\subsection{Total energy of atoms and ions: LDA, GGA, Meta-GGA and SIC}
\label{atomsionsE}

Ground-state energies of neutral atoms from $Z=2$ to $Z=18$ are shown in
Tables \ref{tableatoms1} to \ref{tableatoms3}. In Table \ref{tableatoms1}
we compare selfconsistent LDA(PZ\cite{pz81}) and GGA(PBE\cite{pbe}) 
energies with energies obtained by three approximate schemes: post-LDA 
implementation of GGA and globally and locally scaled selfconsistency.
For global scaling we follow our above discussion and use 
\be
v_{xc}^{GGA(GSSC)}[n]\r=\frac{E_{xc}^{GGA}[n]}{E_{xc}^{LDA}[n]}v_{xc}^{LDA}[n]\r
\ee
as $xc$ potential, throughout the selfconsistency cycle. For local 
scaling, the prefactor of $v_{xc}^{LDA}\r$ is replaced by the
corresponding energy densities. Although GGAs are nowadays often 
implemented fully selfconsistency, application
of approximate schemes to GGA is interesting precisely because fully
selfconsistent data are available for comparison. Table \ref{tableatoms1}
clearly shows that both approximate schemes provide excellent
approximations to the fully selfconsistent values.

\begin{table}[t]
\begin{ruledtabular}
\caption{\label{tableatoms1} Ground-state energies of neutral atoms 
(in Rydberg) obtained selfconsistently with LDA and with GGA, and 
approximate GGA energies predicted by globally (F) and locally (f) scaled 
selfconsistency, starting from LDA, and by post-LDA application (P) of GGA.}
\begin{tabular}{c|c|c|c|c|c}
atom & LDA  & F & f & P & GGA \\ 
\hline
He &  -5.6686  &  -5.7854 & -5.7838 & -5.7844 &  -5.7859  \\ 
Li &  -14.6853 	&  -14.9235 & -14.9206& -14.9220 &  -14.9244\\
Be &  -28.8924  &  -29.2589 & -29.2561& -29.2573 &  -29.2599 \\
B  &  -48.7037  &  -49.2089 & -49.2057& -49.2075 &  -49.2108\\
C  &  -74.9315  &  -75.5846 & -75.5813& -75.5835 &  -75.5874  \\
N  &  -108.258  &  -109.068 & -109.065& -109.067 &  -109.072 \\
O  &  -149.042  &  -149.998 & -149.995& -149.998 &  -150.002\\
F  &  -198.217  &  -199.327 & -199.325& -199.327 &  -199.331\\
Ne &  -256.455  &  -257.729 & -257.726& -257.728 &  -257.733\\
Na &  -322.881  &  -324.341 & -324.338& -324.340 &  -324.345\\
Mg &  -398.265  &  -399.906 & -399.903& -399.905 &  -399.910\\
Al &  -482.628  &  -484.459 & -484.456& -484.458 &  -484.464\\
Si &  -576.429  &  -578.458 & -578.455& -578.458 &  -578.464\\
P  &  -679.991  &  -682.225 & -682.222& -682.224 &  -682.231 \\
S  &  -793.471  &  -795.886 & -795.883& -795.885 &  -795.892\\
Cl &  -917.327  &  -919.935 & -919.932& -919.934 &  -919.941\\
Ar &  -1051.876  &  -1054.686 & -1054.682& -1054.685 &  -1054.692 \\
\end{tabular}
\end{ruledtabular}
\end{table}

Next, we consider two alternative choices of energy functionals, for which 
fully selfconsistent results are not readily available: meta-GGA and
self-interaction corrections. In Table 
\ref{tableatoms2} we compare a GSCC implementation of 
meta-GGA(TPSS)\cite{tpss1,tpss2} with results obtained from a post-GGA(PBE) 
implementation of the same functional. Fully selfconsistent implementation 
of meta-GGA would require the OEP algorithm, and has not yet been reported. 
However, previous tests of meta-GGA suggest that orbital selfconsistency and 
post-GGA implementation of MGGA give rather similar results, and provide 
significant improvement on GGA.\cite{mggatests1,mggatests2} 
Consistently with these observations, and also with those of 
Ref.~\onlinecite{cafiero} (which, however, uses the now obsolete PKZB form of 
meta-GGA), we find, in Table~\ref{tableatoms2}, very close agreement 
between scaled and post-GGA implementations of TPSS meta-GGA. Systematically,
the LSSC energies are a little less negative and the GSCC energies very 
slightly more negative than post-GGA energies.

\begin{table}[t]
\begin{ruledtabular}
\caption{\label{tableatoms2} Ground-state energies of neutral atoms obtained
selfconsistently with GGA(PBE), compared to meta-GGA(TPSS) energies predicted 
by globally (F) and locally (f) scaled selfconsistency, starting from GGA,
and by post-GGA calculations (P).}
\begin{tabular}{c|c|c|c|c}
atom   &  GGA   &  F  & f & P\\
\hline
He &  -5.7859 &  -5.8185 & -5.7625 &-5.8183 \\
Li &  -14.9244 &  -14.9768 & -14.9111 &-14.9767 \\
Be &  -29.2599 &  -29.3414 & -29.2654 &-29.3412 \\
B  &  -49.2108 &  -49.3077 & -49.2061 &-49.3076 \\
C  &  -75.5874 &  -75.7091 & -75.5768 &-75.7089 \\
N  &  -109.0715 &  -109.2304 & -109.0610 &-109.2302 \\
O  &  -150.0019 &  -150.1664 & -149.9781 &-150.1662 \\
F  &  -199.3313 &  -199.5190 & -199.3022 &-199.5188 \\
Ne &  -257.7329 &  -257.9607 & -257.7085 &-257.9605 \\
Na &  -324.3454 &  -324.5951 & -324.3452 &-324.5949 \\
Mg &  -399.9103 &  -400.1831 & -399.9298 &-400.1830 \\
Al &  -484.4641 &  -484.7584 & -484.4871 &-484.7583 \\
Si &  -578.4640 &  -578.7866 & -578.4969 &-578.7865 \\
P  &  -682.2314 &  -682.5997 & -682.2784 &-682.5896 \\
S  &  -795.8921 &  -796.2723 & -795.9499 &-796.2722 \\
Cl &  -919.9413 &  -920.3493 & -920.0101 &-920.3492 \\
Ar &  -1054.6923 & -1055.1359  & -1054.7761 &-1055.1359 \\
\end{tabular}
\end{ruledtabular}
\end{table}

A different type of density functional is represented by the self-interaction 
correction (SIC). Although several distinct such 
corrections have been proposed, we here focus on the best known suggestion, 
made by Perdew and Zunger in 1981 (PZSIC).\cite{pz81} PZSIC provides an 
orbital-dependent correction of the form
\bea
E_{xc}^{approx,SIC}[n_\ua,n_\da]= 
\nonumber \\
E_{xc}^{approx}[n_\ua,n_\da]-\sum_{i,\sigma}
\left(E_H[n_{i\sigma}]-E_{xc}^{approx}[n_{i\sigma},0]\right),
\eea
which can be applied to any approximate density functional $E_{xc}^{approx}$.
Here we chose this functional to be the LDA. The correction terms 
depend on the partial density of each occupied orbital, $n_{i\sigma}\r$,
and not explicitly on the total density. Hence, selfconsistent implementation
should proceed via the OEP.\cite{norman}
Instead, common implementations of PZSIC follow
a suggestion made in the original reference,\cite{pz81} and vary the energy 
functional with respect to the orbitals. The resulting single-particle
equation is not the usual Kohn-Sham equation, but features an effective
potential that is different for each orbital. 

In Table~\ref{tableatoms3} 
we compare energies obtained from solution of this equation with those 
obtained, in a much simpler way, from applying GSCC to LDA+PZSIC. 
Very good agreement is achieved. 
This is even more remarkable as in usual implementations of PZSIC the
potentials are different for each orbital, and the resulting orbitals
are not orthogonal, whereas in the GSSC implementation the local potential
is the same for all orbitals, which are automatically orthogonal. 
Formally, the GSSC implementation is thus more satisfying than the 
standard implementation. 

\begin{table}[t]
\begin{ruledtabular}
\caption{\label{tableatoms3} Ground-state energies of neutral atoms obtained
selfconsistently with LDA, and PZSIC energies predicted by globally scaled 
selfconsistency, starting from LDA, and by post-LDA implementation of PZSIC,
all compared to energies obtained from a standard (i.e., 
orbitally selfconsistent) implementation of LDA+PZSIC.}
\begin{tabular}{c|c|c|c|c}
atom & LDA & F  & P & LDA+PZSIC\\
\hline
He &  -5.6686  & -5.8366  & -5.8329 &-5.8386\\
Li &  -14.6853  &  -15.0075 & -15.0036 &-15.0091 \\
Be &  -28.8924  &  -29.3863 & -29.3824 &-29.3877 \\
B  &  -48.7037  &  -49.3992 & -49.3946 &-49.4007\\
C  &  -74.9315  &  -75.8573 & -75.8520 &-75.8592 \\
N  &  -108.258  &  -109.442 & -109.436 &-109.445\\
O  &  -149.042  &  -150.505 & -150.497 &-150.508\\
F  &  -198.217  &  -199.987 & -199.979 &-199.991\\
Ne &  -256.455  &  -258.561 & -258.551 &-258.565\\
Na &  -322.881  &  -325.332 & -325.324 &-325.336\\
Mg &  -398.265  &  -401.060 & -401.053 &-401.064\\
Al &  -482.628  &  -485.769 & -485.762 &-485.773  \\
Si &  -576.429  &  -579.922 & -579.916 &-579.926\\
P  &  -679.991  &  -683.843 & -683.836 &-683.846\\
S  &  -793.471  &  -797.689 & -797.683 &-797.692\\
Cl &  -917.327  &  -921.920 & -921.914 &-921.923\\
Ar & -1051.876   &  -1056.850 & -1056.850 &-1056.854 \\
\end{tabular}
\end{ruledtabular}
\end{table}

\begin{figure}[t]
\centering
\includegraphics[width=0.4 \textwidth,angle=0]{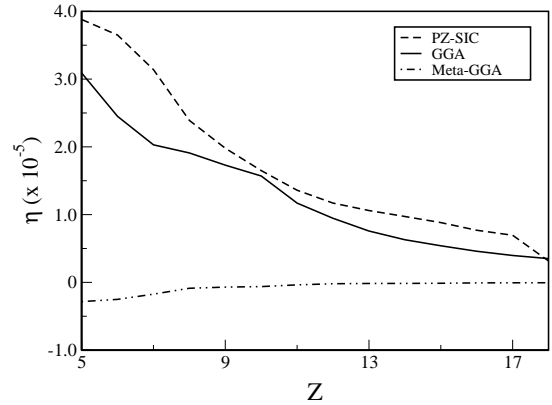}
\caption {\label{atomsfig1} Relative deviation, as defined in 
Eq.~(\ref{relerr}), of GSSC GGA compared to selfconsistent GGA (full line), 
of GSSC meta-GGA compared to post-GGA meta-GGA (dash-dotted line) 
and of GSCC LDA+PZSIC compared to
orbitally selfconsistent LDA+PZSIC. Note that only in the first case this
relative deviation can be interpreted as a relative error of GSSC relative
to fully selfconsistent results, as in the other two cases different
approximations are compared among each other.}
\end{figure}

The analysis of total energies of neutral atoms is summarized in 
Figure~\ref{atomsfig1}, which displays the relative deviation
\be
\eta = \frac{\left(E_{ref}-E_{GSSC}\right)}{E_{ref}}
\label{relerr}
\ee
of GSCC from selfconsistent data (GGA), post-GGA data (meta-GGA) and
orbitally selfconsistent data (PZSIC).

Table \ref{atomerrors} presents a representative error analysis. Values
of $C_2$, defined in Eq.~(\ref{criterium2}), indicate that the term neglected 
in the GSSC approach is smaller than the term kept, but not by a
sufficient margin to explain, on its own, the very good approximation to total
energies obtained from global scaling. The breakdown of the error of the
total energy into the contributions arising from each term individually,
shows that there is very substantial error cancellation, mostly between the
sum of the KS eigenvalues and the potential energy in the $xc$ potential.
As a consequence of this error cancellation, which is ultimately due to the 
variational principle, selfconsistent total energies are predicted by
scaling approaches with much higher accuracy than the values of $C_2$ suggest.

\begin{table}
\caption{\label{atomerrors}
Breakdown of errors of the different contributions to the
total energy in Eq.~(\ref{etot}), comparing LDA-based global-scaling
predictions of GGA energies (GSSC GGA) with selfconsistent GGA energies.
The column labeled $C_2$ contains the integrated validity criterium of
global scaling, as defined in Eq.~(\ref{criterium2}).
Note that we have not defined an integrated validity
criterium for local scaling.}
\begin{ruledtabular}
\begin{tabular}{c|c|c|c|c|c|c}
  atom & $C_2$ & $\Delta E_{KS}$ & $\Delta E_H$ & $\Delta V_{xc}$& $\Delta E_{xc}$ & $\Delta E_0$\\
\hline
He & 0.016 & 0.0974 &0.0131 & 0.0766 & -0.0081 & -0.0004 \\
C  & 0.115 & 0.2684 &-0.1670& 0.4504 & 0.0122  & -0.0028 \\
O  & 0.016 & 0.333  &-0.318 & 0.679  & 0.024   & -0.004 \\
Na & 0.353 & 0.464  &-0.541 & 1.041  & 0.031   & -0.004 \\
Si & 0.147 & 0.689  &-0.691 & 1.416  & 0.030   & -0.006 \\
Ar & 0.009 & 0.959  &-0.948 & 1.950  & 0.036   & -0.007 \\
\end{tabular}
\end{ruledtabular}
\end{table}

\begin{table}[t]
\begin{ruledtabular}
\caption{\label{tableions}Ground-state energies of positive ions,
obtained from self-consistent LDA and GGA calculations, as well as from
locally and globally scaled LDA calculations.}
\begin{tabular}{c|c|c|c|c|c}
atom & LDA   & F & f & P & GGA\\
\hline
He$^+$  &  -3.8838  & -3.9873 & -3.9853 &-3.9864 &  -3.9875\\
Li$^+$  &  -14.2831  & -14.5130 & -14.5105 &-14.5116 &  -14.5135\\
Be$^+$  &  -28.2278  &-28.5977  & -28.5945 &-28.5960 &  -28.5986\\
B$^+$   &  - 48.0742 & -48.5858 & -48.5827 &-48.5840 &  -48.5869\\
C $^+$  &  -74.0698  & -74.7273 & -74.7238 &-74.7258 &  -74.7292\\
N$^+$   &  -107.159  & -107.973 & -107.969 &-107.972 &  -107.976\\
O$^+$   &  -148.014  &-148.993  & -148.990 &-148.992 &  -148.997\\
F$^+$   &  -196.886  & -198.016 & -198.013 &-198.015 &  -198.020\\
Ne$^+$  &  -254.823  & -256.113 & -256.110 &-256.112 &  -256.117\\
Na$^+$  &  -322.487  & -323.947 & -323.944 &-323.946 &  -323.951\\
Mg$^+$  &  -397.696  &-399.346  & -399.344 &-399.346 &  -399.351\\
Al$^+$  &  -482.188  & -484.021 & -484.018 &-484.020 &  -484.026 \\
Si$^+$  &  -575.823  &-577.851  & -577.848 &-577.850 &  -577.856 \\
P$^+$   &  -679.219  &-681.450  & -681.447 &-681.449 &  -681.456\\
S$^+$   &  -792.690  &-795.133  & -795.129 &-795.132 &  -795.139\\
Cl$^+$  &  -916.350  &-918.975  & -918.971 &-918.974 &  -918.981\\
Ar$^+$  &  -1050.703  & -1053.524 & -1053.520 &-1053.523 &  -1053.530\\
\end{tabular}
\end{ruledtabular}
\end{table}

The performance of global and local scaling, as well as that of post
methods, for (positive) ions is essentially the same obtained for
neutral systems, as is illustrated for selected ions in Table~\ref{tableions}.

From the data in Tables~\ref{tableatoms1} to \ref{tableions} we conclude
that global scaling yields slightly better ground-state energies than local 
scaling and post methods. In addition, for orbital-dependent functionals,
such as PZ-SIC, it provides a common local potential and orthogonal orbitals
at no extra computational cost.

\subsection{Kohn-Sham eigenvalues of atoms and ions: LDA, GGA, Meta-GGA and SIC}
\label{atomsionseps}

In principle, a further advantage of scaled approaches, as compared to
post methods, is that the former also provide corrections to the Kohn-Sham 
spectrum, which cannot be obtained from the latter. Instead of considering 
the entire spectrum, we focus on the highest occupied eigenvalue, as it is 
physically most significant. Representative data for this eigenvalue 
are collected in Tables~\ref{tableatoms5} to \ref{tableatoms8}.

\begin{table}[t]
\begin{ruledtabular}
\caption{\label{tableatoms5} Highest occupied KS eigenvalues of neutral
atoms, obtained selfconsistently from LDA and from GGA, and LDA-based
predictions of the GGA energies by means of global and local scaling.}
\begin{tabular}{c|c|c|c|c}
atom & LDA  & F & f & GGA \\
\hline
He &  -1.1404   &  -1.2073  &  -1.2187  &  -1.1585\\   
Li &  -0.2326  &  -0.2558 &  -0.2548 &  -0.2372\\
Be &  -0.4120  &  -0.4455 &  -0.4328 &  -0.4122\\
B  &  -0.2997  &  -0.3387 &  -0.3368 &  -0.2955\\
C  &  -0.4517  &  -0.5008 &  -0.5013 &  -0.4482\\
N  &  -0.6137  &  -0.6716 &  -0.6760 &  -0.6103 \\
O  &  -0.5502  &  -0.5951 &  -0.6082 &  -0.5287\\
F  &  -0.7701  &  -0.8242 &  -0.8387 &  -0.7523\\
Ne &  -0.9955  &  -1.0576  &  -1.0757  &  -0.9810\\
Na &  -0.2263  &  -0.2427 &  -0.2515 &  -0.2234  \\
Mg &  -0.3513  &  -0.3724 &  -0.3703 &  -0.3454\\
Al &  -0.2212  &  -0.2417 &  -0.2455 &  -0.2218 \\
Si &  -0.3383  &  -0.3648 &  -0.3707 &  -0.3403 \\
P  &  -0.4602  &  -0.4922 &  -0.5007 &  -0.4626\\
S  &  -0.4612  &  -0.4898 &  -0.4971 &  -0.4403 \\
Cl &  -0.6113  &  -0.6454 &  -0.6519 &  -0.5978\\
Ar &  -0.7646  &  -0.8039 &  -0.8114 &  -0.7560\\
\end{tabular}
\end{ruledtabular}
\end{table}

\begin{table}[t]
\begin{ruledtabular}
\caption{\label{tableatoms6} Highest occupied KS eigenvalues of positive ions,
obtained selfconsistently from LDA and from GGA, and LDA-based
predictions of the GGA energies by means of global and local scaling.}
\begin{tabular}{c|c|c|c|c}
atom & LDA   & F & f & GGA\\
\hline
He$^+$ &  -3.0211  &  -3.1623 & -3.1610 &  -3.0898 \\
Li$^+$ &  -4.3793  &  -4.5188 & -4.5348 &  -4.4400\\
Be$^+$ &  -1.0509  &  -1.0969 & -1.0954 &  -1.0637 \\
B$^+$  &  -1.4267  &  -1.4827 & -1.4663 &  -1.4336\\
C$^+$  &  -1.3072  &  -1.3724 & -1.3702 &  -1.3031\\
N$^+$  &  -1.6213  &  -1.6973 & -1.6968 &  -1.6197\\
O$^+$  &  -1.9384  &  -2.0233 & -2.0269 &  -1.9381\\
F$^+$  &  -1.9307  &  -1.9973 & -2.0140 &  -1.9091\\
Ne$^+$ &  -2.3093  &  -2.3860 & -2.4043 &  -2.2926\\
Na$^+$ &  -2.6857  &  -2.7709 & -2.7932 &  -2.6734 \\
Mg$^+$ &  -0.8803 &  -0.9072 & -0.9252 &  -0.8746 \\
Al$^+$ &  -1.0949  &  -1.1257  & -1.1278  &  -1.0867 \\
Si$^+$ &  -0.8904 &  -0.9224 & -0.9284 &  -0.8912 \\
P$^+$  &  -1.1005  &  -1.1387  & -1.1461  &  -1.1046\\
S$^+$  &  -1.3115  &  -1.3556  & -1.3651  &  -1.3172 \\
Cl$^+$ &  -1.3592  &  -1.3983  & -1.4075  &  -1.3376 \\
Ar$^+$ &  -1.5975  &  -1.6426  & -1.6510  &  -1.5851\\
\end{tabular}
\end{ruledtabular}
\end{table}

\begin{table}[t]
\begin{ruledtabular}
\caption{\label{tableatoms7} Highest occupied KS eigenvalues of neutral
atoms, obtained from GGA and predicted for TPSS meta-GGA by global and
local scaling.}
\begin{tabular}{c|c|c|c}
atom   &  GGA   &   F  &  f \\
\hline
He &  -1.1585  &  -1.1767  & -0.7159\\
Li &  -0.2372 &  -0.2420 & -0.1055\\
Be &  -0.4122 &  -0.4192 & -0.2249\\
B  &  -0.2955 &  -0.3023 & -0.0968\\
C  &  -0.4482 &  -0.4566 & -0.1763\\
N  &  -0.6103 &  -0.6208 & -0.2656\\
O  &  -0.5287 &  -0.5356 & -0.2585\\
F  &  -0.7523 &  -0.7606 & -0.3881\\
Ne &  -0.9810 &  -0.9912 &-0.5245\\
Na &  -0.2234 &  -0.2259 &-0.1000\\
Mg &  -0.3454 &  -0.3486 &-0.1841\\
Al &  -0.2218 &  -0.2249 &-0.0731\\
Si &  -0.3403 &  -0.3443 &-0.1364\\
P  &  -0.4626 &  -0.4675 &-0.2065\\
S  &  -0.4403 &  -0.4444 &-0.2202\\
Cl &  -0.5978 &  -0.6028 &-0.3154\\
Ar &  -0.7560 &  -0.7618 & -0.4140 \\
\end{tabular}
\end{ruledtabular}
\end{table}

\begin{table}[t]
\begin{ruledtabular}
\caption{\label{tableatoms8} Highest occupied KS eigenvalues of neutral
atoms, obtained selfconsistently from LDA and predicted for LDA+PZSIC by 
means of global scaling, compared to an orbitally selfconsistent implementation
of LDA+PZSIC.}
\begin{tabular}{c|c|c|c}
atom & LDA  & F  & LDA+PZSIC\\
\hline
He &  -1.1404   &  -1.2370  &  -1.8957\\
Li &  -0.2326  &  -0.2642 &  -0.3927\\
Be &  -0.4120  &  -0.4575 &  -0.6554\\
B  &  -0.2997  &  -0.3538 &  -0.6125\\
C  &  -0.4517  &  -0.5219 &  -0.8512\\
N  &  -0.6137  &  -0.6990 &  -1.0960\\
O  &  -0.5502  &  -0.6195 &  -1.0627 \\
F  &  -0.7701  &  -0.8570 &  -1.3732 \\
Ne &  -0.9955  &  -1.0991  &  -1.6838 \\
Na &  -0.2263  &  -0.2544 &  -0.3782 \\
Mg &  -0.3513  &  -0.3875 &  -0.5502\\
Al &  -0.2212  &  -0.2571 &  -0.4081\\
Si &  -0.3383  &  -0.3844 &  -0.5717 \\
P  &  -0.4602  &  -0.5159 &  -0.7376 \\
S  &  -0.4612  &  -0.5116 &  -0.7696\\
Cl &  -0.6113  &  -0.6718 &  -0.9632 \\
Ar &  -0.7646  &  -0.8348 &  -1.1586 \\
\end{tabular}
\end{ruledtabular}
\end{table}

Unlike total energies, KS eigenvalues (single-particle energies) are not
protected by a variational principle, and simple approximations, such as
global or local scaling, may work less well than for total energies. 
For the transition from LDA to GGA, the data in Tables~\ref{tableatoms5}
and \ref{tableatoms6} show indeed that not applying any scaling factor at 
all, i.e., using the uncorrected LDA eigenvalues, produces better 
approximations to the GGA eigenvalues than either local or global scaling.
For meta-GGA (Table~\ref{tableatoms7}), no fully selfconsistent eigenvalues 
are available for comparison (as explained in the introduction, this would 
require the OEP algorithm to be implemented for meta-GGA). For PZ-SIC, on 
the other hand, Table~\ref{tableatoms8} shows that
substantial improvement on the LDA eigenvalues is obtained by global
scaling, although not by the same margin observed for total energies.

\section{Tests and applications to models of extended systems}
\label{modeltests} 

The calculations in the preceding sections were restricted to atoms and ions.
Some results for molecules were already reported in Ref.~\onlinecite{cafiero}.
We therefore next turn to models for extended systems.

\subsection{Hubbard model}
\label{hubbard}

First, we consider the Hubbard model, which is a much studied model Hamiltonian
of condensed-matter physics, for which the basic theorems of density-functional 
theory all hold.\cite{gs,lsoc,coldatoms}
In the present context, this model constitutes a most interesting
test case for scaled selfconsistency for three different reasons:
(i) It is maximally different from the atoms and ions considered in the
previous sections, and thus provides tests in an entirely different
region of functional space. (ii) For a small number of lattice sites 
the {\em exact} diagonalization of the Hamiltonian matrix in a complete
basis (consisting of one orbital per site) can be performed numerically,
hence providing exact data against which all approximate functionals and
implementations can be checked. (iii) In the thermodynamic limit (infinite 
number of sites) with equal occupation on each sites (homogeneous density 
distribution) the {\em exact} many-body solution of the one-dimensional 
Hubbard model is known from the Bethe-Ansatz technique.\cite{liebwu} This
solution allows the construction of the exact local-density approximation,
\cite{gs,coldatoms} which circumvents the need for analytical 
parametrizations
of the underlying uniform reference data, required for the conventional LDA
of the {\em ab initio} Coulomb Hamiltonian. The simultaneous disponibility 
of the exact solution and the exact LDA makes this model an ideal test case
for DFT approximations.

The one-dimensional Hubbard model is specified by the second-quantized 
Hamiltonian
\bea
\hat{H}=
-t\sum^L_{i,\sigma} (c_{i\sigma}^\dagger c_{i+1,\sigma}+H.c.)
+U\sum^L_i c_{i\ua}^\dagger c_{i\ua}c_{i\da}^\dagger c_{i\da}
\label{hm}
\eea
defined on a chain of $L$ sites $i$, with one orbital per site.
Here $U$ parametrizes the on-site interaction and $t$ the hopping between
neighbouring sites. Below all energies and values of $U$ are given in
multiples of $t$, as is common practice in studies of the model (\ref{hm}).

\begin{table}
\caption{\label{hubbardtable1}
Exact per-site ground-state energy (to six significant digits), selfconsistent 
LDA energy, selfconsistent Hartree energy, and Hartree-based simulations of 
the LDA energy via global scaling, local scaling and post-Hartree
implementation. All energies have been multiplied by $-10/t$.
First set of three rows: $L=10$ sites with $N=2$ electrons.
Second set of three rows: $L=10$ sites with $N=8$ electrons.
Third set of three rows: $L=100$ sites with $N=96$ electrons.
All calculations were done for open boundary conditions with $v_{ext}=0$.}
\begin{ruledtabular}
\begin{tabular}{c|c|c|c|c|c|c|c}
$N$ & $U$ & exact & Hartree  & F & f & P & LDA\\
\hline
2 & 2 & 3.69905  & 3.58412  & 3.68371 & 3.68457 & 3.68456 & 3.68472\\
  & 4 & 3.65957  & 3.35102  & 3.62921 & 3.63175 & 3.63014 & 3.63239\\
  & 6 & 3.64244  & 3.12796  & 3.60332 & 3.60710 & 3.60117 & 3.60815\\
\hline
8 & 2 & 8.87176 & 8.24130  & 8.81004 & 8.81062 & 8.81064 & 8.81067\\
  & 4 & 7.30440  & 5.02351   & 7.13510 & 7.13573 & 7.13529 & 7.13574\\
  & 6 & 6.37228  & 1.81375  & 6.11910 & 6.11850 & 6.11520 & 6.11918\\
\hline
96& 2 &- & 8.02648& 8.71172 & 8.71174 & 8.71174  & 8.71174 \\
  & 4 &- & 3.41821 & 6.19811 & 6.19806 &  6.19791 & 6.19962 \\
  & 6 &-& -1.18994  & 4.74410 & 4.74411 & 4.74369 & 4.75018\\
\end{tabular}
\end{ruledtabular}
\end{table}

The availability of an exact many-body solution for small $L$, and
of the exact LDA, permit us to eliminate many of the uncertainties 
associated with more approximate calculations.
To test the ideas of scaled selfconsistency for the Hubbard model, we 
attempt to predict the energies of a selfconsistent LDA calculation by
starting with a simple Hartree (mean-field) calculation. Equation
(\ref{gssc}) cannot be directly applied to the $xc$ potentials because
the $xc$ potential of a pure Hartree calculation is zero, but we can
apply it to the entire interaction-dependent contribution to the effective 
potential, {\em i.e.}, to the sum of Hartree and $xc$ terms. We thus
approximate the entire interaction potential by its Hartree contribution, 
and use scaled selfconsistency to predict the values of a Hartree+LDA
calculation. In analogy to our previous equations we write this as
\be
v_{int}^{GSSC(H+LDA)}(i)=\frac{E_{int}^{H+LDA}[n_i]}{E_{int}^H[n_i]}v^H(i).
\ee
Results obtained from this approximation can be compared to a
selfconsistent Hartree+LDA calculation, in which
\be
v_{int}(i)=v^H(i)+v_{xc}^{LDA}(i).
\ee
This comparison provides a severe test for the scaled selfconsistency 
concept, as the starting functional (Hartree) is quite different from the 
target functional (Hartree+LDA).

The effective potential $v_s(i)$ is, in principle, given by adding 
$v_{int}(i)$ to the external potential $v_{ext}(i)$, but here we chose 
$v_{ext}(i)\equiv 0$, so that the interaction-dependent contribution becomes 
identical to $v_s(i)$. This makes the test even tougher, as there is no large 
external potential dominating the total energy and the eigenvalues, and 
potentially masking effects of $v_{xc}^{LDA}(i)$.

The system becomes inhomogeneous because of the finite size of the chain.
For $L=10$ sites exact energies can still be obtained, and are displayed,
together with various approximations to them, in Table~\ref{hubbardtable1}.
For $L=100$ sites obtaining exact energies is out of question, but all
DFT procedures are still easily applicable. Corresponding results for
total energies are also displayed in Table~\ref{hubbardtable1}.
Eigenvalues are recorded in Table~\ref{hubbardtable2}. 

\begin{table}
\caption{\label{hubbardtable2}Highest occupied Kohn-Sham eigenvalue
obtained from selfconsistent LDA calculations, selfconsistent Hartree 
calculations, and Hartree-based simulations of the LDA energy via global 
scaling (F) and local scaling (f). The systems are the same as in 
Table~\ref{hubbardtable1}.}
\begin{ruledtabular}
\begin{tabular}{c|c|c|c|c|c}
$N$ & $U$ & Hartree  & F & f  & LDA\\
\hline
2 & 2  & -1.67177  & -1.76773 & -1.76999 & -1.74131 \\
  & 4  &  -1.44839  & -1.71514 & -1.72036  &  -1.66731\\
  & 6  & -1.23452  & -1.69022 & -1.69737 & -1.63056 \\
\hline
8 & 2  & -0.02259  &  -0.16428 &  -0.16486 & -0.09917\\
  & 4  &  0.77959 & 0.25307 &  0.25211   &  0.39835\\
  & 6  & 1.58054  & 0.50636 & 0.50562 & 0.68490\\ 
\hline
96& 2  & 0.80452  & 0.66180 & 0.66175 & 0.74151 \\
  & 4  &  1.76435  &  1.18533 & 1.18531 & 1.25131\\
  & 6  &  2.72424 & 1.48818 & 1.48819 & 1.47135 \\
\end{tabular}
\end{ruledtabular}
\end{table}

From Tables~\ref{hubbardtable1} and \ref{hubbardtable2} we conclude that:
(i) For ground-state energies the LDA typically deviates by about 1\% from 
the exact values, while a Hartree calculation is about 10\% off. 
(ii) Taking the selfconsistent LDA data as standard, we find that for 
ground-state energies locally scaled selfconsistency comes closest,
followed, in this order, by post-Hartree data, globally scaled 
selfconsistency and the original Hartree calculation. 
(iii) For eigenvalues the order changes: globally scaled selfconsistency
does better than locally scaled selfconsistency, whereas post-calculations
naturally do not provide any correction at all, and come a distant third.

\begin{table}
\caption{\label{hubbardtable3} 
Validity criterium and error analysis for the system
of Tables~\ref{hubbardtable1} and \ref{hubbardtable2}.
$C_2$ is defined in Eq.~(\ref{criterium2}), and the other columns report,
for each term of the total energy expression (\ref{etothub}), the
difference between the fully selfconsistent result and the result
obtained by GSSC. The error in the total energy is given by
$\Delta E_0=\Delta E_{KS} - \Delta V_{int}+\Delta E_{int}$, and
always significantly lower than each of the individual errors.
First set of nine lines: global scaling. Second set of nine lines:
local scaling. Recall that we have not defined an integrated validity
criterium for local scaling.} 
\begin{ruledtabular}
\begin{tabular}{c|c|c|c|c|c|c}
$N$ & $U$ & $C_2$ & $\Delta E_{KS}$ & $\Delta V_{int}$ & $\Delta E_{int}$
& $\Delta E_0$\\
\hline
2 & 2 &  0.162 & 0.05285 & 0.04794 & -0.00592 & -0.00101 \\
  & 4 & 0.247 & 0.09565 & 0.08495 & -0.01388 & -0.00318 \\
  & 6 & 0.290 & 0.11933 & 0.10522 & -0.01894 & -0.00483 \\
\hline
8 & 2 & 0.098 & 0.52085 & 0.51601 & -0.00547 & -0.00063 \\
  & 4 & 0.135 & 1.16243 & 1.15674 & -0.00633 & -0.00064 \\
  & 6 &  0.135 & 1.42893 & 1.42831 & -0.00070 & -0.00008 \\
\hline
96& 2 &  0.098 & 7.6541 & 7.6531  & -0.0012  & -0.0002 \\
  & 4 &  0.065 & 6.4775 & 6.4813 & -0.0113 & -0.0151 \\
  & 6 &  0.018 & -0.7711 & -0.7835 & -0.0731 & -0.0608 \\
\hline
\hline
2 & 2 &   -    & 0.05737 & 0.05526 & -0.00227 & -0.00016 \\
  & 4 &   -    & 0.10610 & 0.10078 & -0.00596 & -0.00064 \\
  & 6 &   -    & 0.13363 & 0.12627 & -0.00841 & -0.00105 \\
\hline
8 & 2 &   -    & 0.52543 & 0.52400 & -0.00148 & -0.00005 \\
  & 4 &   -    & 1.17013 & 1.17090 & 0.00075 & -0.00002 \\
  & 6 &   -    & 1.43479 & 1.44124 & 0.00577 & -0.00068 \\
\hline
96& 2 & - & 7.6551 & 7.6548  & -0.0003 & 0.0000 \\
  & 4 & - & 6.4768 & 6.4811 & -0.0114 & -0.0156 \\
  & 6 & - & -0.7709 & -0.7834 & -0.0731 & -0.0607 \\
\end{tabular}
\end{ruledtabular}
\end{table}

We have performed similar comparisons also for other systems (different 
number of sites $L$, different number of electrons $N$, different values of
the on-site interaction parameter $U$), but the general trend is the same,
although in isolated cases the relative quality of F, f and P-type
implementations can be different.

In Table~\ref{hubbardtable3}  we analyse the satisfaction 
of the validity criteria of Sec.~\ref{validity} and the contribution of
each term in the ground-state-energy expression to the total error. The 
values of $C_2$ in Table~\ref{hubbardtable3} show that the 
term neglected in global scaling is always smaller than the term kept, 
but in unfavorable cases, such as low density and strong interactions, 
can be a sizeable fraction of it. Hence, just as for atoms and ions, 
additional error cancellation,
arising from the separate contributions to the ground-state energy,
is taking place in these situations.

These contributions to the ground-state energy can be written as 
\bea
E_0=\sum_i\epsilon_i - \int d^3r \, n\r v_{int}[n]\r + E_{int}[n]
\\
=:E_{KS} - V_{int} + E_{int},
\label{etothub}
\eea
and their errors are recorded in Table \ref{hubbardtable3}. 

The data in Table~\ref{hubbardtable3} show that the total error is much 
less than that of each contribution individually. Global and local SSC 
thus benefit from systematic and extensive error compensation 
-- mostly between the sum of the KS eigenvalues and the interaction
potential energy -- which make it applicable even when the simplest from 
of the validity criterium, Eq.~(\ref{criterium1}), or its integrated
version, Eq.~(\ref{criterium2}),
are violated. Clearly, locally scaled selfconsistence benefits even 
more than globally scaled selfconsistence.

\subsection{Quantum wells}
\label{qwell}

Moving up from one dimension to three, we next consider semiconductor
heterostructures. To illustrate the main features of scaled
selfconsistency we consider a simple quantum well, in which the electrons 
are free to move along the $x$ and $y$ direction, but confined along the
$z$ direction. The modelling of such quantum wells by means of the
effective-mass approximation within DFT is described in Ref. 
\onlinecite{qwell1} and the particular approach we use is that of Ref. 
\onlinecite{qwell2}, whose treatment we follow closely, and to which we 
refer the reader for more details.

In Table \ref{welltable1} we display ground-state energies and 
highest occupied KS eigenvalue obtained for representative quantum wells by 
means of a selfconsistent LDA calculation (using the PW92\cite{pw92} 
parametrization for the electron-liquid correlation energy), a selfconsistent 
GGA calculation (using the PBE\cite{pbe} form of the GGA), post-LDA GGA, 
and globally-scaled and locally-scaled simulations of PBE by means of LDA.

\begin{table}
\caption{\label{welltable1} 
Ground-state energies (in $meV$ per particle) and highest-occupied KS
eigenvalue (in $meV$) of a quantum well of depth $200 meV$, width $10
nm$, embedded in a background semiconductor (charge reservoir) of
width $50 nm$, areal density of $n_A=10^{12} cm^{-2}$, effective
electron mass of $0.1m_0$ and relative dielectric constant
$\varepsilon=10$. These parameters are typical of semiconductor
heterostructures.\cite{qwell3,qwell4} For the second set of two lines
we have added a central barrier of width $3nm$ and hight $200 meV$,
dividing the system in two weakly coupled halves.}
\begin{ruledtabular}
\begin{tabular}{r|c|c|c|c|c}
           & LDA(PW92)    &    F    &     f   & P &   GGA(PBE)\\
\hline
E          & 78.0197   & 77.7135 & 77.7194 & 77.7136 & 77.7122\\
$\epsilon$ & 117.455   & 117.060 & 116.993 & 117.455 & 117.183 \\
\hline
E          & 129.134  & 128.435 & 128.443 & 128.436  & 128.426 \\
$\epsilon$ & 163.322  & 162.420 & 162.375 & 163.322  & 162.706 \\
\end{tabular}
\end{ruledtabular}
\end{table}

Total energy differences between all three approximate schemes are
marginal, compared to the difference between a pure LDA calculation
and a GGA calculation: all three LDA-based schemes closely approximate the
results of selfconsistent GGA calculations, with global scaling slightly 
better than the other two. Interestingly, this is the opposite trend observed 
in the Hubbard chain, where local scaling was best. For eigenvalues,
post-calculations do not provide any improvement, whereas both global and
local scaling come close to the fully selfconsistent results.

In Table \ref{welltable2} we show the values of $C_2$, and the breakdown of
the error of the ground-state energy in its components, according to
\bea
E_0[n]=\sum_i\epsilon_i - E_H[n] -\int d^3r \, n\r v_{xc}\r + E_{xc}[n]
\\
=E_{KS} - E_H[n] - V_{xc} + E_{xc}[n].
\eea

The GSSC validity criterium (\ref{criterium2}) is well satisfied.
Values of $C_2$ are an order of magnitude smaller than for the Hubbard
chain. This reduction expresses the fact that simulating a GGA by starting 
from an LDA is a much easier task than simulating an LDA by starting 
from a Hartree calculation. Also understandable are the slightly larger 
values of $C_2$ and of the errors in the energy found for the double well 
as compared to the single well, since this structure has more pronounced 
density gradients, enhancing the difference between LDA and GGA, and 
complicating the task the scaling factor has to accomplish. 

For the energy components, we observe the same type of substantial error 
cancellation found also in the calculations for atoms, ions and the Hubbard 
model, resulting 
in very good total energies. We note that this error cancellation is
taking place mainly between the sum of the KS eigenvalues and the potential
energy in the $xc$ potential, whose errors are subtracted in forming $E_0$.
The errors in the $xc$ and the Hartree energies are orders of magnitude 
smaller. This is the same observation previously made for the other systems, 
suggesting that this pattern of error cancellation is a general trend.

\begin{table}
\caption{\label{welltable2} 
Error analysis for the heterostructures of Table~\ref{welltable1}. First
set of two lines: simple well.  Second set of two lines: well with
barrier. The total error is obtained as $\Delta E_0=\Delta
E_{KS}-\Delta E_H-\Delta V_{xc}+\Delta E_{xc}$. Very substantial error
cancellation between the sum of the KS eigenvalues and the $xc$
potential energy is taking place.}
\begin{ruledtabular}
\begin{tabular}{c|c|c|c|c|c|c}
  & $C_2$ & $\Delta E_{KS}$ & $\Delta E_H$ & $\Delta V_{xc}$& $\Delta E_{xc}$ & $\Delta E_0$\\
\hline
GSSC  & 0.022 & -0.1232 & -0.0129 & -0.1056 & 0.0061 & 0.0013 \\
LSSC  &   -   & -0.1904 & -0.0953 & -0.0703 & 0.0321 & 0.0072 \\
\hline
GSSC  & 0.036 & -0.237  & 0.036  & -0.248 & 0.035 & 0.010 \\
LSSC  &   -   & -0.289 & -0.032 & -0.216 & 0.058 & 0.017
\end{tabular}
\end{ruledtabular}
\end{table}

\section{Conclusions}
\label{concl}

Three different possibilities for approximating the results of selfconsistent 
calculations with a complicated functional by means of selfconsistent
calculations with a simpler functional have been compared: Post-selfconsistent
implementation of the complicated functional (P), global scaling of
the potential of the simple functional by the ratio of the energies (F), and
local scaling of the potential of the simple functional by the ratio of the 
energy densities (f). While method P is a standard procedure of DFT, 
method F was proposed only very recently (and without a solid derivation or
a validity criterium, both of which we provide here). Method f is proposed
here for the first time.

These three approaches were tested for atoms, ions, Hubbard chains and
quantum wells, and applied to Hartree, LDA, GGA, meta-GGA and SIC
type density functionals. Our conclusions are summarized as follows.

(i) All three prescriptions provide significant improvements on the total
ground-state energies obtained from the simple functional (as measured
by their proximity to those obtained selfconsistently from the complex
functional). For these energies, on average, globally scaled selfconsistency 
has a slight advantage compared to post-selfconsistent implementations and 
local scaling, but the relative performance of all three methods depends on 
fine details of the system parameters, on the quantity calculated, and on the 
particular density functional used. In general, neither local scaling nor 
alternative scaling schemes (employing other scaling factors, or scaling other 
contributions to the energy) provide consistent improvements on global scaling,
which for total energies is, on average, the best of all methods tested here.

(ii) An additional advantage of scaled selfconsistency is that it also provides
approximations to the eigenvalues, eigenfunctions and effective potentials
of the complicated functional, which is by construction impossible for
post-methods. Indeed, for orbital-dependent potentials, such as PZ-SIC, scaling 
provides a simple and effective way to produce a common local potential and 
orthogonal orbitals. For models of extended systems, scaling provides 
significant improvement on eigenvalues obtained from the simple functional, 
although by a smaller margin than for the total energy. For finite systems,
similar improvement was found in applications of PZ-SIC, but not in tests 
trying to predict GGA eigenvalues from LDA.

(iii) The reason scaled selfconsistency works is due to an interplay of
three distinct mechanisms. First, whenever the term neglected in 
Eq.~(\ref{identity}) is much smaller than the term kept, its neglect is 
obviously a good approximation. This is checked by the point-wise criterium
(\ref{criterium1}). Second, even when the neglected term is not much smaller,
quantities that depend on the potential at all points in space, such as the
total energies or the eigenvalues, benefit from error cancellation arising 
from different points in space, as described by the integrated criterium
(\ref{criterium2}). Total energies --- but not eigenvalues --- also
benefit from an additional error cancellation between the different
contributions to the total energy expression of DFT. Numerically, we found
this third mechanism to be dominant. The fact that this
additional error cancellation operates only for total energies explains why 
in all our tests these are consistently better described by scaled
selfconsistency than eigenvalues.

Different scaling schemes from the two employed here may be investigated, and 
certainly additional information from application to still other classes of
systems should be useful, but it seems safe to conclude from the present
analysis that scaled selfconsistency is a most useful concept for
density-functional theory, allowing the efficient and reliable 
implementation of density functionals of hitherto unprecedented complexity,
without ever requiring their variational derivative with respect to
either orbitals or densities. 

{\bf Acknowledgments}
This work was sup\-por\-ted by FAPESP and CNPq.

\end{document}